# NONLINEAR RESONANCES IN THE SOLAR SYSTEM

Renu Malhotra
Lunar and Planetary Institute
3600 Bay Area Boulevard, Houston, TX
Tel: (713)486-2114; Fax: (713)486-2162
E-mail: renu@lpi.jsc.nasa.gov

**Abstract**

Orbital resonances are ubiquitous in the Solar system. They play a decisive role in the long term dynamics, and in some cases the physical evolution, of the planets and of their natural satellites, as well as the evolution of small bodies (including dust) in the planetary system. The few-body gravitational problem of hierarchical planetary-type systems allows for a complex range of dynamical timescales, from the fast orbital periods to the very slow orbit precession rates. The interaction of fast and slow degrees of freedom produces a rich diversity of resonance phenomena. Weak dissipative effects — such as tides or radiation drag forces — also produce unexpectedly rich dynamical behaviors. This paper provides a mostly qualitative discussion of simple dynamical models for the commonly encountered orbital resonance phenomena in the Solar system.

**Key words**: *planetary system, orbits, resonances, resonance passage, resonance overlap*

# 1. Introduction

Soon after Newton's formulation of the Universal Law of Gravitation it became evident that a dynamical model of the Solar System based upon a simple superposition of two-body motions was not adequate for the observations. The mutual interactions of the planets (and of their natural satellites) were necessary to fit the observations and infer various properties of the system. Theoretical efforts to determine the effects of planetary interactions led to the development of perturbation theory and averaging methods, and to many other developments in mathematics. In recent years, celestial mechanics has evolved from its traditional habitat in the study of orbits of particular Solar System objects to the study of the structure of extended regions of phase space (of planetary systems, of satellite systems, and of other small celestial bodies), as well as of evolution due to dissipative effects. The greatest difficulties in these studies — and also the most interesting dynamical behaviors — are associated with orbital resonances.

Even in the simplest system consisting of only two planets, there are six degrees of freedom corresponding to the three spatial degrees of freedom for each planet. Therefore, what at first glance would appear to be only a two-frequency system (i.e., the frequencies of revolution of the two planets around the Sun) is actually one with six frequencies. Two of these are the obvious ones of revolution around the Sun; the other four (two for each planet) are the much slower frequencies of precession of the *orientation* of the orbits. The existence of both fast and slow degrees of freedom produces a rich diversity of orbital resonance phenomena.

As a general classification, there are two types of orbital resonances. The intuitively most obvious type, referred to simply as "orbital resonance" or "mean-motion resonance", occurs when the orbital periods of two planets – or satellites – are nearly commensurate. The second type, called "secular resonance", involves commensurabilities between the slow frequencies of precession of the orientation of orbits. Classical perturbation theory runs into the notorious problem of "small divisors" in analyzing the mutual gravitational perturbations of two planets near a resonance. In the case of the first type of resonance, a significant but subtle difficulty is that there are actually *several* resonances in the vicinity of any orbital period commensurability. This "resonance splitting" arises due to the precession of the orientation of orbits.

The multiplicity of resonances is a consequence of the coupling between perturbations of different timescales, and, in general, may produce chaotic behavior in the vicinity of an orbital resonance. However, in many cases, the behavior at a resonance is largely regular, albeit complicated by non-linear effects. Examples include the observed orbital resonances amongst the satellites of Jupiter and Saturn, where "single resonance theory" works [1], and many phenomena in planetary rings where resonant perturbations from satellites are implicated in a bewildering variety of features such as gaps, kinks and sharp edges, and the confinement of narrow rings [2, 3]. In other cases, interactions between neighboring resonances become important; these lead to a variety of different phenomena associated with secondary resonances and chaotic dynamics. A well-known example is the "Kirkwood gaps" in the asteroid belt, which are deficits in the number distribution of asteroids at several locations corresponding to mean motion resonances with Jupiter; chaos due to interacting resonances has been identified as a mechanism for producing at least one of these gaps [4, 5, 6] and for transporting meteorites from the main asteroid belt to the Earth [7, 8]. Chaotic resonances are of great significance in the formation of planetary systems [9, 10] – in planetesimal dynamics, the transport of planetesimals to a proto-planet, the interactions of proto-planets, and finally the clearing of planetesimal debris from interplanetary space [11, 12, 13]. An interesting and very complicated example is that of Pluto which exhibits several secular resonances in addition to a 3:2 orbital period resonance with Neptune [14]. Secular resonances may play a determining role in the long term dynamical stability of a planetary system [15, 16].

Dissipative effects (such as gas drag and mass loss in the Solar Nebula, tides in the case of planetary satellites, collisions in planetary rings, and radiation forces in the case of dust particles) cause orbits to evolve across resonances, leading to separatrix-crossing phenomena and capture into resonance. Pluto's very peculiar orbit (it crosses the orbit of Neptune but is dynamically protected from close encounters by the resonance phase locking) may owe its origin to orbit evolution of the outer planets due to mass loss in the late stages of planet formation [17]. The preponderance of orbital resonance locks amongst the Jovian and Saturnian satellites is thought to be due to the differential orbit evolution induced by tidal dissipation [1, 18, 19, 20, 21, 22]. These satellites probably did not form in resonant orbits; however, once established, the orbital phase locking can be maintained for long times. Similar evolution probably occurred in the Uranian satellite system also. However, owing to a small but crucial difference in parameters, the orbital resonances in this system were not long-lived [23, 24]. Uranus is less oblate than Jupiter and Saturn; in addition, the relative masses of the satellites and their orbital radii conspire to cause an insufficient splitting of resonances; as a result, the interactions between neighboring resonances give rise to an instability of resonant orbits. Tidal evolution and passage through resonance can have a profound effect on the geophysical evolution of satellites [20, 22, 23, 24, 25, 26, 27]. Orbital resonance phenomena are also found in the dynamics of circumplanetary and interplanetary dust particles that are perturbed by radiation and electromagnetic forces. For example, the dusty ring around Jupiter exhibits features that are attributed to "Lorentz resonances" (commensurabilities between the Keplerian orbit period and the spin period of the planetary magnetic field) that dominate the orbital evolution of charged dust grains [28]. As a final example, I can mention the recent prediction and detection of a "Solar ring" of asteroidal dust particles that spiral inward due to Poynting-Robertson light drag, and are captured into long-lived orbital resonances with the Earth [29, 30].

The diversity of orbital resonance dynamics precludes a complete treatment in these few short pages. Therefore, this paper is confined to giving a brief review of dynamical models for the simplest of the commonly encountered orbital resonance problems in the Solar System. No attempt has been made to give exhaustive citations to the literature, but the selected references should provide the interested reader with sufficient leads to the technical literature.

## 2. Basic formalism

A peculiarity of the unperturbed two-body bounded motion is its complete degeneracy (see, for example, [31]): the motion is periodic in all three spatial degrees of freedom, and the three frequencies are all equal (to the orbital frequency). This property allows a choice of several different coordinates in which the motion is separable. The variables that have been used traditionally to characterize the Keplerian motion are the classical orbital elements (see Figure 1)[1]:

$a$     semimajor axis
$e$     eccentricity
$i$     inclination
$\Omega$     longitude of ascending node
$\varpi$     longitude of periapse
$\lambda$     mean longitude

These elements form a complete set equivalent to the six Cartesian components of the relative position and velocity vectors, $\mathbf{r} = \mathbf{R}_1 - \mathbf{R}_0$ and $\mathbf{v} = \mathbf{V}_1 - \mathbf{V}_0$, in the two-body system. The coordinates and velocities can be transformed by means of a canonical transformation to action-angle variables (see, for example, section 143 in [33]); these are given as follows in terms of the orbital elements:

$$\begin{aligned} J_1 &= \mu\sqrt{G\mathcal{M}a}, & \theta_1 &= \lambda \\ J_2 &= J_1(1 - \sqrt{1-e^2}), & \theta_2 &= -\varpi \\ J_3 &= J_1(1-e^2)^{1/2}(1-\cos i), & \theta_3 &= -\Omega \end{aligned} \quad (1)$$

where $\mu = m_0 m_1 / \mathcal{M}$ is the reduced mass, and $\mathcal{M} = m_0 + m_1$ is the total mass of the two-body system consisting of the primary, $m_0$ and the satellite, $m_1$.[2] For a system with $N > 1$ satellites about the primary, one can define $N$ sets of canonical variables in a coordinate system found by Jacobi (see, for example, p. 539-540 in [34]). These coordinates are mathematically equivalent to the set $(\boldsymbol{J}, \boldsymbol{\theta})$ for a hierarchy of $N$ two-body systems (but of course, the $(\boldsymbol{J}, \boldsymbol{\theta})$ are no longer action-angle variables when the mutual interactions of the satellites are taken into account). The $j$-th two-body system consists of a total mass $\mathcal{M}_j = m_0 + m_1 + ... + m_j$ and a reduced mass, $\mu_j = m_j(\mathcal{M}_j - m_j)/\mathcal{M}_j$, with relative position and velocity vectors, $\mathbf{r}_j = \mathbf{R}_j - \mathbf{R}_{\text{cm}}^{(j-1)}$ and $\mathbf{v}_j = \mathbf{V}_j - \mathbf{V}_{\text{cm}}^{(j-1)}$, where $(\mathbf{R}_j, \mathbf{V}_j)$ and $(\mathbf{R}_{\text{cm}}^{(j-1)}, \mathbf{V}_{\text{cm}}^{(j-1)})$ are the position and velocity of $m_j$ and of the center of mass of $m_0, m_1, ..., m_{j-1}$, respectively.

---
[1] The definitions of $\varpi$ and $\lambda$ may appear counter-intuitive, but arise naturally in celestial mechanics calculations. For a more complete description and motivation for this choice of variables, the reader may consult any celestial mechanics text, for example, chapter 6 in [32].

[2] In the context of the Solar System, the central body (either the Sun or a planet) is always much more massive than the secondaries (the planets or satellites, respectively). We use the term 'primary' for the massive central body and 'satellite' for the secondaries.

The three-body problem, with a dominant central body and two smaller bodies in nearly circular, co-planar orbits, is a fundamental model for perturbation analysis in Solar System dynamical studies. The Hamiltonian for the planetary three-body problem, truncated to the second order in the masses, is given by

$$H = H_0 + H_{12}, \quad (2)$$

$$H_0 = -\frac{Gm_0 m_1}{2a_1} - \frac{Gm_0 m_2}{2a_2} \quad (3)$$

$$H_{12} = -Gm_1 m_2 \left\{ \frac{1}{|\mathbf{r}_1 - \mathbf{r}_2|} - \frac{\mathbf{r}_1 \cdot \mathbf{r}_2}{r_2^3} \right\} \quad (4)$$

The unperturbed part, $H_0$, is the sum of a pair of two-body Hamiltonians, and $H_{12}$ is the interaction potential.

In the case of planetary satellite systems, an important additional perturbation is that due to the oblateness of the planet; the orbit-averaged oblateness perturbation potential is given by (cf. section 11-3(C) in [35])

$$H_{\text{ob}} = -\sum_j \frac{Gm_0 m_j}{4a_j} \mathcal{J}_2 \left(\frac{R}{a_j}\right)^2 (1 - e_j^2)^{-3/2} (3\cos^2 i_j - 1), \quad (5)$$

where $R$ and $\mathcal{J}_2$ are the equatorial radius and oblateness of the planet.

The perturbation, $H_{12}$, can be expanded in a Fourier cosine series in the angular variables, $\{\lambda_j, \varpi_j, \Omega_j\}$, with coefficients that are power series in the usually small orbital eccentricities and inclinations. This expansion takes the following form [34, 36]

$$H_{12} \simeq -\frac{Gm_1 m_2}{a_2} \sum_{k,q} S_{k,q} \cos \phi_{k,q} \quad (6)$$

where
$$\phi_{k,q} = (k+q)\lambda_2 - k\lambda_1 - (q_1 \varpi_1 + q_2 \varpi_2 + q_3 \Omega_1 + q_4 \Omega_2), \qquad q = q_1 + q_2 + q_3 + q_4 \quad (7)$$

and, to leading order in $e$ and $i$,
$$S_{k,q} \simeq f_{k,q}(\alpha) e_1^{|q_1|} e_2^{|q_2|} i_1^{|q_3|} i_2^{|q_4|} \quad (8)$$

where $\alpha = a_1/a_2 < 1$ is the ratio of the semimajor axes of the two orbits, and $f_{k,q}(\alpha)$ is a function of Laplace coefficients [34, 36]. The sum is taken over all integer values of $k$ and $q_i$. Note that each term of the series satisfies the so-called d'Alembert characteristic: the integer coefficient of $\varpi$ or $\Omega$ in the argument of the cosine corresponds to the power of $e$ (resp. $i$) in the coefficient $S$. This property permits a choice of non-singular canonical variables for the analysis (see Section 3 and Eqn. 12 below). A further condition on the integer coefficients, $q_3 + q_4 = $ even, is also realized; this condition means that the lowest order terms involving the inclinations are of the second order.

In practice, in most cases of interest only a few lowest order terms in the above series are used to analyze the dynamics. Of special note are the "secular" terms – those whose argument $\phi_{k,q}$ is independent of the mean longitudes, $\lambda_j$, and depends only upon the slow angles $\varpi_j$ and $\Omega_j$. For orbital configurations that are far from low-order resonances, the secular terms alone (together with $H_{\text{ob}}$) provide an adequate description of the long-term behavior of the system: the semimajor axes remain unperturbed, while the shape and orientation of the orbits exhibit long-period variations, including a secular precession of the nodes and periapses.

The secular motions of the nodes and periapses are important in separating the frequencies of several angular arguments $\phi_{k,q}$ having the same ratio of coefficients of the mean longitudes, $\lambda_1$ and $\lambda_2$.

An orbital resonance exists if the relative orbital motion of two satellites is period. This happens when the orbital frequencies, $n_j = \dot{\lambda}_j$, are commensurate, i.e., the ratio $n_2/n_1$ is close to a ratio of small integers, $p : p + q$. The integer $q$ is called the order of the resonance. In a resonant configuration, the longitude of the satellites at every $q$-th conjunction (i.e., when they pass each other) is nearly fixed or oscillates about a slowly varying center. Stable oscillation of successive conjunctions can be maintained only about certain well-defined longitudes. A qualitative feature of a stable resonance configuration is that conjunctions of the satellites occur away from the longitude at which the orbits are closest together. The possible centers of oscillation of conjunctions can be uniquely specified with respect to the periapses, $\varpi_j$, or nodes, $\Omega_j$.

The periodic repetition of the geometry of the system enhances the effects of the mutual gravitational interaction of the satellites, so that over many successive conjunctions, their small mutual perturbations can accumulate into large (but usually bounded) changes of the orbital parameters. What follows below is a description of the conditions for exact resonance and the multiplicity of a $p : p + q$ orbital resonance.

Sufficiently close to a specific $p : p+q$ resonance, the frequencies of some of the terms in Eqn. 6 (those with $k = p$) become very small. For $q = 1$ or $q = 2$, to second order in $e_1, e_2, i_1$ and $i_2$, there are six terms with very low frequencies associated with any orbital commensurability; each of these defines a unique resonance condition. (For larger values of $q$, there is an even greater multiplicity of the split resonances.) As illustration, consider the 2:1 and 3:1 commensurabilities. Table I lists the defining resonance condition, the corresponding resonance angle $\phi_{p,q}$ (which oscillates when the resonance condition is satisfied, but rotates otherwise), and a nomenclature for these resonances. The latter is derived from the coefficient of the leading resonant term in the perturbing potential. The subscripts 1 and 2 in the table refer to the inner and outer orbits, respectively.

**Table I: Splitting of resonances**

| resonance condition | resonance angle | 'name' of the resonance |
|---|---|---|
| $\langle 4\dot{\lambda}_2 - 2\dot{\lambda}_1 - 2\dot{\Omega}_2 \rangle = 0$ | $4\lambda_2 - 2\lambda_1 - 2\Omega_2$ | 4:2 $i_2^2$ resonance |
| $\langle 4\dot{\lambda}_2 - 2\dot{\lambda}_1 - \dot{\Omega}_2 - \dot{\Omega}_1 \rangle = 0$ | $4\lambda_2 - 2\lambda_1 - \Omega_2 - \Omega_1$ | 4:2 $i_1 i_2$ resonance |
| $\langle 4\dot{\lambda}_2 - 2\dot{\lambda}_1 - 2\dot{\Omega}_1 \rangle = 0$ | $4\lambda_2 - 2\lambda_1 - 2\Omega_1$ | 4:2 $i_1^2$ resonance |
| $\langle 2\dot{\lambda}_2 - \dot{\lambda}_1 - \dot{\varpi}_1 \rangle = 0$ | $2\lambda_2 - \lambda_1 - \varpi_1$ | 2:1 $e_1$ resonance |
| $\langle 4\dot{\lambda}_2 - 2\dot{\lambda}_1 - \dot{\varpi}_2 - \dot{\varpi}_1 \rangle = 0$ | $4\lambda_2 - 2\lambda_1 - \varpi_2 - \varpi_1$ | 4:2 $e_1 e_2$ resonance |
| $\langle 2\dot{\lambda}_2 - \dot{\lambda}_1 - \dot{\varpi}_2 \rangle = 0$ | $2\lambda_2 - \lambda_1 - \varpi_2$ | 2:1 $e_2$ resonance |
| $\langle 3\dot{\lambda}_2 - \dot{\lambda}_1 - 2\dot{\Omega}_2 \rangle = 0$ | $3\lambda_2 - \lambda_1 - 2\Omega_2$ | 3:1 $i_2^2$ resonance |
| $\langle 3\dot{\lambda}_2 - \dot{\lambda}_1 - \dot{\Omega}_2 - \dot{\Omega}_1 \rangle = 0$ | $3\lambda_2 - \lambda_1 - \Omega_2 - \Omega_1$ | 3:1 $i_1 i_2$ resonance |
| $\langle 3\dot{\lambda}_2 - \dot{\lambda}_1 - 2\dot{\Omega}_1 \rangle = 0$ | $3\lambda_2 - \lambda_1 - 2\Omega_1$ | 3:1 $i_1^2$ resonance |
| $\langle 3\dot{\lambda}_2 - \dot{\lambda}_1 - 2\dot{\varpi}_1 \rangle = 0$ | $3\lambda_2 - \lambda_1 - 2\varpi_1$ | 3:1 $e_1^2$ resonance |
| $\langle 3\dot{\lambda}_2 - \dot{\lambda}_1 - \dot{\varpi}_2 - \dot{\varpi}_1 \rangle = 0$ | $3\lambda_2 - \lambda_1 - \varpi_2 - \varpi_1$ | 3:1 $e_1 e_2$ resonance |
| $\langle 3\dot{\lambda}_2 - \dot{\lambda}_1 - 2\dot{\varpi}_2 \rangle = 0$ | $3\lambda_2 - \lambda_1 - 2\varpi_2$ | 3:1 $e_2^2$ resonance |

The form of the resonance angles suggests a linear transformation to "resonance variables" via the generating function:

$$\begin{aligned}
\mathcal{F} &= I_1\Big((p+q)\lambda_2 - p\lambda_1 - q\Omega_1\Big)/q + I_2\Big((p+q)\lambda_2 - p\lambda_1 - q\Omega_2\Big)/q \\
&\quad + I_3\Big((p+q)\lambda_2 - p\lambda_1 - q\varpi_1\Big)/q + I_4\Big((p+q)\lambda_2 - p\lambda_1 - q\varpi_2\Big)/q \\
&\quad + \Gamma_1 \lambda_1 + \Gamma_2 \lambda_2
\end{aligned} \quad (9)$$

Using the definitions of the canonical variables in Eqns. 1, this transformation yields the following resonance variables:

$$\phi_1 = \Big((p+q)\lambda_2 - p\lambda_1 - q\Omega_1\Big)/q, \quad I_1 = \mu_1 \sqrt{G\mathcal{M}_1 a_1 (1 - e_1^2)}(1 - \cos i_1) \simeq \frac{1}{2} m_1 \sqrt{G m_0 a_1}\, i_1^2$$

$$\phi_2 = \Big((p+q)\lambda_2 - p\lambda_1 - q\Omega_2\Big)/q, \quad I_2 = \mu_2 \sqrt{G\mathcal{M}_2 a_2 (1 - e_2^2)}(1 - \cos i_2) \simeq \frac{1}{2} m_2 \sqrt{G m_0 a_2}\, i_2^2$$

$$\begin{aligned}
\phi_3 &= \big((p+q)\lambda_2 - p\lambda_1 - q\varpi_1\big)/q, & I_3 &= \mu_1\sqrt{G\mathcal{M}_1 a_1}\Big(1 - \sqrt{1-e_1^2}\Big) \simeq \frac{1}{2}m_1\sqrt{Gm_0 a_1}\,e_1^2 \\
\phi_4 &= \big((p+q)\lambda_2 - p\lambda_1 - q\varpi_2\big)/q, & I_4 &= \mu_2\sqrt{G\mathcal{M}_2 a_2}\Big(1 - \sqrt{1-e_2^2}\Big) \simeq \frac{1}{2}m_2\sqrt{Gm_0 a_2}\,e_2^2 \\
\gamma_1 &= \lambda_1, & \Gamma_1 &= \mu_1\sqrt{G\mathcal{M}_1 a_1} + \frac{p}{q}(I_1 + I_2 + I_3 + I_4) \qquad (10)\\
\gamma_2 &= \lambda_2, & \Gamma_2 &= \mu_2\sqrt{G\mathcal{M}_2 a_2} - \frac{p+q}{q}(I_1 + I_2 + I_3 + I_4)
\end{aligned}$$

Averaging $H$ over the non-resonant, "fast" variables, $\gamma_1$ and $\gamma_2$, one gets a Hamiltonian with 4 degrees of freedom $(\phi_1, ..., \phi_4)$; $\Gamma_1$ and $\Gamma_2$ are first integrals for the averaged Hamiltonian. Referring to Table I, it can be seen that the $\phi_1$ and $\phi_2$ degrees of freedom are coupled together by the 'mixed' $i_1 i_2$ resonance; similarly for the $\phi_3$ and $\phi_4$ degrees of freedom. Furthermore, to the second order in $e_j$ and $i_j$, there are no resonant couplings across the eccentricity-type and the inclination-type resonances. Therefore, in a further simplification, one can consider a resonance Hamiltonian with only two degrees of freedom: $(\phi_1, I_1; \phi_2, I_2)$ for the inclination-type (or "vertical") resonances, or $(\phi_3, I_3; \phi_4, I_4)$ for the eccentricity-type (or "horizontal") resonances.

The resonances are *well separated* if the rates $\dot{\varpi}_j, \dot{\Omega}_j$ are sufficiently different from each other that the resonance splitting is larger than the widths of neighboring resonances. In such cases, "single resonance theory" provides a good description of the resonance dynamics. In many other cases of interest, the degeneracy is only imperfectly removed, and the interactions amongst two or more neighboring resonances lead to secondary resonances and chaotic layers. These different types of behavior are discussed in the following sections.

### 3. Single resonance theory

In the simplest case of resonance, only one of the resonance angles is truly a slow variable. Then, one can once more average over the faster angles and set to constants the corresponding conjugate momenta; this reduces the number of degrees of freedom to one. The single resonance Hamiltonian has the general form, $K(I, \phi) = K_0(I) + V(I, \phi)$. As the most common cases of interest in planetary and satellite dynamics involve small-to-modest orbital eccentricities and inclinations, it is sufficient to use a second-order Taylor series expansion of $K_0(I)$, and retain only the leading term in $V(I, \phi)$. The symmetry properties of the planetary three-body system yield the following single resonance Hamiltonian:

$$K(I, \phi) = \nu I - \beta I^2 + \mu(2I)^{q/2} \cos q\phi. \qquad (11)$$

$K(I, \phi)$ is a pendulum-like Hamiltonian. $\nu$ is a measure of the "distance" of the satellites from the exact resonance condition for unperturbed circular orbits; $\beta$ measures the nonlinearity and $\mu$ the strength of the resonant interaction between the satellites[3]. Figures 2 and 3 show the phase space in Poincare variables,

$$x = \sqrt{2I}\cos\phi \qquad \text{and} \qquad y = \sqrt{2I}\sin\phi, \qquad (12)$$

for the case $q = 1$ (first order resonance), and $q = 2$ (second order resonance).

The topology of the phase space is as follows. All phase space orbits are periodic, but a critical trajectory, the *separatrix*, whose period is unbounded, exists for $\nu$ greater than a critical value $\nu_c$ given by

$$\nu_c = \begin{cases} 3|\mu^2\beta/4|^{1/3} & \text{for } q = 1 \\ 2|\mu| & \text{for } q = 2 \end{cases} \qquad (13)$$

When it exists, the separatrix may have one or two branches; it divides the phase plane into two or three zones: an *external* zone with circulating orbits outside the outer branch of the separatrix; an *internal* zone inside the inner branch of the separatrix (when there are two branches); and a *resonance* zone between the two branches of the separatrix (when there are two branches) or inside the separatrix (when there is only one branch). Except for a narrow range of parameters and initial conditions, most orbits in the resonance zone are *librating* orbits, i.e. the resonance angle, $\phi$, executes finite amplitude oscillations, whereas most orbits in the external and internal zones are circulating orbits ($\phi$ increases or decreases without bound). The half-width of the resonance zone is approximately,

$$\Delta I \simeq \left[\frac{2\mu}{\beta}(2I_r)^{q/2}\right]^{1/2} \qquad (14)$$

where $I_r \simeq \nu/2\beta$ is the value at the center of the resonance zone; $\Delta I$ defines the maximum deviation of the canonical momentum, $I$, from its exact-resonance value for which the resonance angle can still librate.

---

[3] A slight notational degeneracy: the $\mu$ here should not be confused with the two-body reduced mass defined in the previous section.

**Adiabatic evolution**

The principal effects of energy dissipation or external torques on satellite orbits are to evolve the orbital semimajor axes, and thus to make the parameter $\nu$ change with time. For sufficiently slow variation of $\nu$, the evolution of the system (Eqn. 11) can be evaluated by means of the adiabatic theorem (e.g. section 49 in [31]) and a few simple ideas regarding separatrix-crossing transitions. The reader may consult references [1, 37, 38, 39, 40] for detailed explanations; here we give only the highlights of the theory.

The satellite system describes a trajectory that is coincident with a closed ("guiding") trajectory in the $(x, y)$ phase plane of the frozen system with the instantaneous value of $\nu$. As long as $\nu$ does not change very much during the period of the guiding trajectory, the action is an adiabatic invariant; the action is proportional to the area enclosed by the guiding trajectory:

$$J = \frac{1}{2\pi} \oint I \, d\phi = \frac{1}{2\pi} \oint x \, dy \qquad (15)$$

As $\nu$ changes slowly, the phase space of the satellite system evolves. The guiding trajectory also evolves, slowly changing shape while preserving the enclosed area. The exception to the adiabatic invariance of $J$ occurs when the guiding trajectory becomes nearly coincident with the separatrix. In this situation, the period of the guiding trajectory becomes very large, and hence the change in $\nu$ in one period becomes increasingly large, violating the adiabatic assumption. As $\nu$ continues to change, the guiding trajectory must cross the separatrix. During this transition, while the system is in the neighborhood of the separatrix, an instantaneous action integral is not well defined. Nevertheless, in the subsequent evolution, the separatrix will evolve away from the guiding trajectory, and the new action will once again be an adiabatic invariant.

In order to apply this theory to orbital resonances, it is useful to recall the following.

- $\nu$ (which measures the deviation from exact commensurability) is an increasing or decreasing function of time according as the ratio of the orbital radii, $\alpha = a_1/a_2$, is an increasing or decreasing function of time due to the external torques.

- The areas enclosed in the *resonance* zone and in the *internal* zone (see Figs. 2 and 3) increase with $\nu$. Also, the distance of the center of the resonance zone from the origin is an increasing function of $\nu$.

- From the definitions (Eqns. 10), $\sqrt{2I}$ is proportional to the orbital eccentricity (or inclination). Therefore, the *average* radius of a trajectory in the $(x, y)$ phase plane is proportional to the *average* eccentricity (or inclination).

- For circulating trajectories (in the *internal* or *external* zones), the action, $J$, is also proportional to the average eccentricity (or inclination).

- For trajectories in the resonance zone, $J$ measures the libration amplitude; the distance of the center of the resonance zone from the origin is proportional to the average eccentricity (or inclination).

The adiabatic theory for passage through resonance can be applied directly to determine the evolution of the orbital parameters. As an example, consider the evolution across an $e_1$-resonance. Let the initial value of the eccentricity be $e_{1,0}$ and corresponding action $J_0$. Away from the separatrix, $J_0$ is an adiabatic invariant as $\nu$ changes slowly. The values of the action on the internal and external branches of the separatrix are functions of $\nu$, denoted by $J_{\text{int}}(\nu)$ and $J_{\text{ext}}(\nu)$, respectively. There are two cases to consider. In following the analysis below, it is useful to refer to the phase portraits in Figures 2 and 3.

- *$\nu/\nu_c$ decreasing from a large positive value* (i.e. the satellites approach the resonance on 'diverging' orbits): Initially the resonance zone is located a large distance from the origin. Assuming that the guiding trajectory is initially close to the origin, it will lie in the *internal* zone whose total area decreases with time. At some value, $\nu = \nu_T$, when the internal zone shrinks sufficiently that $J_{\text{int}}(\nu_T) \simeq J_0$, the guiding trajectory makes a transition across the separatrix to the *external* zone. (A transition into the resonance zone is not possible as its area is decreasing with time.) The new value of the action, $J_f$, is the value of the action on the outer branch of the separatrix at $\nu = \nu_T$, i.e. $J_f \simeq J_{\text{ext}}(\nu_T)$. Therefore, upon crossing the resonance, the eccentricity "jumps" to a larger value.

- *$\nu/\nu_c$ increasing from a large negative value* (i.e. the satellites approach the resonance on 'converging' orbits): In this case, initially there is no separatrix, and the guiding trajectory is a circulating trajectory. At $\nu_c$ a separatrix appears with values of the action, $J_{\text{int}}(\nu_c) = 0$, and $J_{\text{ext}}(\nu_c) > 0$. If $J_0 < J_{\text{ext}}(\nu_c)$, then the guiding trajectory evolves smoothly into the *resonance* zone. On the other hand, if $J_0 > J_{\text{ext}}(\nu_c)$, then a separatrix-crossing transition will occur at some value $\nu_T$ greater than $\nu_c$ when the guiding trajectory becomes coincident with the external branch of the separatrix. $\nu_T$ is given implicitly by the equation $J_{\text{ext}}(\nu_T) = J_0$. In this case, transition may occur either into the *resonance* zone or into the *internal* zone because the areas enclosed by both zones are increasing with time. A probability of resonance capture can be calculated as a function of the initial action, $J_0$, or equivalently, as a function of $e_{1,0}$ [38, 40].

In summary, one finds that capture into resonance (libration) can occur only when the satellites approach the resonance on converging orbits (increasing $\nu$). Resonance capture is certain if the initial eccentricity is smaller than a critical value; for larger initial eccentricities, the probability of capture decreases. If capture does occur, then in the subsequent evolution the eccentricity increases secularly as the adiabatic invariance of the action "drags" the satellites along the resonance zone which moves away from the origin as $\nu$ increases. If capture does not occur, then the new trajectory will lie in the *internal* zone, with a final value of the action smaller than its initial value; i.e. the final eccentricity will be smaller than its initial value.

The above analysis has been applied in modeling and inferring the tidal evolution of the satellites of Jupiter and Saturn [1], and also in a theory for the origin of Pluto's resonant orbit [17]. Other interesting applications are in the evolution of dust particles due to electromagnetic and radiation forces in various environments (the rings of the outer planets [28, 41], dust particles in circumsolar orbits[30]), and dusty circumstellar disks [42]), and the evolution of planetesimals and protoplanets in the gaseous environment of the early Solar system [43, 44, 45].

## 4. Interacting resonances

We have seen that for any orbital period commensurability there are actually several possible resonances. In the best of circumstances, these resonances are well separated due to the secular motions of the apses and nodes, and one can analyze each resonance in isolation. The next level of complication arises from the interaction of two neighboring resonances.

The simplest model for this interaction is obtained by treating the coupling between two resonances as a perturbation on the single resonance model (cf. Eqn. 11):

$$K = \nu I - \beta I^2 + \mu(2I)^{q/2} \cos q\phi + \varepsilon\sqrt{2I} \cos(\phi - \Omega t). \tag{16}$$

The form of the perturbation term introduced here is suggested by the coupling terms in the interaction Hamiltonian, Eqn. 6 (see references [24, 46]). The frequency $\Omega$ is approximately the frequency difference between the neighboring resonances. If $\Omega$ is sufficiently large, the single resonance theory can be recovered by the averaging principle. However, as $I$ increases, two properties of this system conspire to cause a breakdown of the single resonance theory: (i) the strength of the perturbation increases, and (ii) the strength of the primary resonance, and therefore its libration frequency, also increases, thus reducing the frequency gap between the neighboring resonances. This results in a chaotic broadening of the separatrix of the primary resonance, as well as the appearance of secondary resonances deep inside the resonance zone of the single resonance model. Secondary resonances occur at commensurabilities between the libration frequency and the rotation frequency of neighboring resonances. An example of this situation is shown in a surface of section in Figure 4.

For initial conditions $(I, \phi)$ inside the resonance libration zone, the Fourier decomposition of the perturbation term is given by

$$\sqrt{2I} \cos(\phi - \Omega t) \sim \sum_{k=-\infty}^{+\infty} A_k \cos((k\omega - \Omega)t + \delta) \qquad (17)$$

where $\omega$ is the unperturbed libration frequency of $\phi$; the amplitudes $A_k$ are exponentially small for large $|k|$. One can see from this that even when $\Omega \gg \omega$, close to the unperturbed separatrix, the secondary resonance condition, $\omega = \Omega/k$, will be satisfied for sufficiently large $k$. In fact, for $|k|$ greater than some $k_{\min}$, all the secondary resonances will overlap and broaden the separatrix into a chaotic layer. The width of the chaotic separatrix is exponentially small with the ratio $\Omega/\omega$.

When $\Omega$ is not too much greater than $\omega$, low-order, isolated secondary resonances can appear near the center of the resonance zone (cf. Fig. 4). Single resonance theory can be constructed for these secondary resonances, with Hamiltonians of the same form as Eqn. 11 [46, 47].

**Adiabatic evolution**

The appearance of secondary resonances near the center of the primary resonance zone leads to a new phenomenon in the tidal evolution of satellite resonances: an orbital resonance between two satellites can be disrupted by means of capture into a secondary resonance. This mechanism is described below.

Resonance capture at a small value of $\langle I \rangle$ is described well by the single resonance theory as the chaotic layer near the separatrix is initially exceedingly small. Upon capture into the resonance, $\langle I \rangle$ increases while the libration amplitude becomes quite small. At several points in the subsequent evolution, as $\langle I \rangle$ continues to increase, the system evolves across several secondary resonances. These resonances are born at the center of the primary resonance and migrate out toward the separatrix as $\langle I \rangle$ increases. Capture into any one of secondary resonances drags the satellites towards the chaotic separatrix and eventually allows an escape from the primary resonance. A schematic diagram of this mechanism is shown in Figure 5.

An example of such interacting resonances occurs in the Uranian satellite system. At present these satellites do not exhibit any orbital resonances, but temporary resonance capture in the past may help explain their inferred thermal history. Furthermore, it has been shown that a temporary 3:1 resonance between Miranda and Umbriel may account for the anomalously high inclination of Miranda [23, 24]. Figure 6 shows a numerical simulation of the tidal evolution through the Miranda-Umbriel 3:1 inclination-type resonances in which the satellites escape from the resonance by the secondary-resonance-capture mechanism. Miranda's temporary residence in a similar eccentricity-type resonance in the past may also help account for its geologically diverse surface, because a high orbital eccentricity would enhance the tidal heating within this icy satellite.

**5. Chaotic resonances**

The interaction of two or more resonances at the same orbital period commensurability produces chaotic layers at the separatrices of the individual resonances. When the frequency separation of the primary resonances is modest, the origin of the chaos is readily interpreted as the accumulation of secondary resonances near the separatrix (cf. Figs. 4 and 5).

An instability of circular orbits near the 3:1 Jovian resonance in the asteroid belt arises due to the interaction of the $e_1^2$ and $e_1 e_2$ resonances, where $e_1$ is the eccentricity of an asteroid and $e_2$ is Jupiter's eccentricity [4, 5, 6, 48]. Many previous investigations that assumed a circular orbit of Jupiter had failed to find any instabilities because Jupiter's eccentricity is essential to the origin of the chaotic zone at this resonance. The depletion of asteroids at some of the Jovian resonance locations is attributed to the chaotic excitation of high orbital eccentricity. Asteroids in highly eccentric orbits become Mars- or even Earth-crossers. Close encounters with these planets then clear out the 3:1 Kirkwood gap. It is natural to conjecture that this mechanism — chaotic dynamics in the planar, elliptic, restricted three-body system of the Sun+Jupiter+asteroid, together with close encounters with the terrestrial planets — may provide a universal explanation for all the Kirkwood gaps. The hopes for this have not been fully realized as yet because of significant differences in the phase space structure at other resonances (notably at the 2:1 Jovian resonance) [48, 49]; a complete solution awaits future efforts. In order to solve this problem, it may be necessary to consider dynamical models more complex than that based upon the planar, elliptic, restricted three-body problem [6, 48, 49].

A large chaotic zone near the separatrix of a resonance may occur even in the case where the perturbing planet is on a circular orbit (i.e., in the planar, circular, restricted three-body problem) if the resonant orbit is sufficiently close to that of the planet. The origin of this chaos lies in the interactions between distinct mean motion resonances, for example, a 3:2 resonance and a 4:3 resonance. An example of this phenomenon is seen in a simple model for the Neptune-Pluto 3:2 resonance [14]. Pluto is modeled as a massless test particle (its mass is only two-thousandths that of Neptune), and Neptune's orbit is assumed circular (Neptune's eccentricity is currently only $\sim 0.01$, while Pluto's is 0.25). Pluto's inclination is also neglected in this model. It is found that the region of stable resonance librations has an amplitude of about 100 degrees and is surrounded by a chaotic zone (Figure 7). The observed libration amplitude of Pluto's known orbit is about 80 degrees, safely inside the stable zone.

Evolution due to weak dissipation across chaotic resonances is a pressing and largely unsolved problem. A handful of numerical studies on the tidal evolution of satellites across chaotic layers may be found in the literature [23, 24, 27], but only a few analytical results are available [50]. It is unclear how to quantify *adiabatic* evolution in this case, for a separation of the timescales of "chaotic diffusion" and of dissipation becomes difficult. For the same reason, numerical modeling of these types of problems is also fraught with difficulties.

## 6. Concluding remarks

The Solar System as a dynamical system is a source of a very diverse and large number of dynamics problems. Many of these have parallels in other areas of physics. This paper gives only a very brief glimpse of the simplest examples of these problems. The analytical approach taken here describes some of the basic mechanisms that operate in the dynamical evolution of planetary systems. Entirely omitted in this paper is a discussion of numerical modeling in celestial mechanics which is a very active area of research; indeed it exemplifies in many ways the state-of-the-art in computational precision available today. The interested reader may consult [51] and references therein. The very large parameter space and very large range of dynamical timescales in the planetary few body problem (with or without weak dissipation) continue to challenge analytical techniques as well as the fastest numerical modeling hardware and software.

## Acknowledgements

I am grateful to David Frenkel for many suggestions that vastly improved the manuscript. I also thank Joe Burns, Stan Peale and an anonymous referee for their helpful comments. This research was done while the author was a Staff Scientist at the Lunar and Planetary Institute which is operated by the Universities Space Research Association under contract no. NASW-4574 with the National Aeronautics and Space Administration. This paper is Lunar and Planetary Institute Contribution no. 831.

**FIGURE CAPTIONS**

**Figure 1:**
This shows the two-body Keplerian orbit in space and the definitions of the orbital elements. $\Omega$ is the *longitude of ascending node*, $\omega$ is called the *argument of periapse*, and $f$ is the *true longitude*. In the set of 'classical' orbital elements, the longitude of periapse is defined as $\varpi = \omega + \Omega$, and the mean longitude $\lambda = \int n\, dt + \varpi$, where $n$ is the orbital frequency. For a circular orbit, $\int n\, dt = f$, but for an elliptical orbit, it is an "averaged" value of the true longitude.

**Figure 2:**
The phase space in Poincaré variables $(x, y)$ for an isolated first-order resonance for (a) $\nu < \nu_c$, (b) $\nu = \nu_c$, and (c) $\nu > \nu_c$. The orbits shown in (d) are the same as those in (c), but are displayed in the $(I, \phi)$ plane; the pendulum-like structure is apparent in these variables.

**Figure 3:**
Same as Figure 2, but for a second-order resonance.

**Figure 4:**
This is a surface of section for the perturbed resonance Hamiltonian (Eqn. 16) for a weak perturbation of frequency near three times the small amplitude libration frequency of the primary resonance. The apparition of secondary resonances and the chaotic broadening of the separatrix is evident.

**Figure 5:**
A schematic picture of the adiabatic passage through an orbital resonance. The horizontal axis is $\alpha = a_1/a_2 < 1$, the ratio of the semimajor axes of the two orbits. Resonance capture is possible if the satellites approach the resonance 'from below', i.e. with $\alpha$ increasing with time. Capture in the orbital resonance is followed by a secular increase in the mean value of the canonical momentum, $I$; capture in a secondary resonance "drags" the orbit out towards the separatrix, allowing escape from the orbital resonance.

**Figure 6:**
A numerical simulation of the tidal evolution of Miranda through a 3:1 inclination-type resonance with Umbriel. Miranda's inclination is pumped up to a high value before capture in a 3/1 secondary resonance disrupts the orbital commensurability. (Reproduced from [24].)

**Figure 7:**
Surface of section in the circular, planar, restricted three-body model for the Neptune-Pluto 3:2 resonance. The variables are $\phi = 3\lambda_P - 2\lambda_N - \varpi_P$, and $J = \sqrt{a_P}(1 - \sqrt{1 - e_P^2})$. (The subscripts $P$ and $N$ refer to Pluto and Neptune, resp.; $a_P$ is measured in units of Neptune's orbital radius which is assumed to be constant.) (Reproduced from [14].)